\documentclass{llncs}
\usepackage{graphicx} 
\usepackage[hyphens]{url}
\usepackage[utf8]{inputenc}
\usepackage[english]{babel}
\usepackage{amsmath}
\usepackage{amsfonts}
\usepackage{amssymb}
\usepackage{booktabs}
\usepackage[amsmath,thref,thmmarks]{ntheorem}
\usepackage[algoruled]{algorithm2e}
\usepackage{algpseudocode}
\usepackage{listings}
\usepackage{tabularx}
\usepackage{tabu}

\begin{document}

\mainmatter       
%

\title{How Relevant is the Long Tail?}
\subtitle{A Relevance Assessment Study on Million Short}

\author{
Philipp Schaer\inst{1} \and 
Philipp Mayr\inst{2} \and \\
Sebastian Sünkler\inst{3} \and 
Dirk Lewandowski\inst{3} 
}
\institute{
Cologne University of Applied Sciences, Cologne, Germany
\and
GESIS -- Leibniz Institute for the Social Sciences, Cologne, Germany\\
\and
Hamburg University of Applied Sciences, Hamburg, Germany\\
}

\date{\today}

\maketitle

\begin{abstract}
Users of web search engines are known to mostly focus on the top ranked results of the search engine result page. While many studies support this well known information seeking pattern only few studies concentrate on the question what users are missing by neglecting lower ranked results. To learn more about the relevance distributions in the so-called long tail we conducted a relevance assessment study with the Million Short long-tail web search engine. While we see a clear difference in the content between the head and the tail of the search engine result list we see no statistical significant differences in the binary relevance judgments and weak significant differences when using graded relevance. The tail contains different but still valuable results. We argue that the long tail can be a rich source for the diversification of web search engine result lists but it needs more evaluation  to clearly describe the differences.
\end{abstract}


\section{Introduction}
The ranking algorithms of web search engines try to predict the relevance of web pages to a given search query by a user. By incorporating hundreds of ``signals" modern search engines do a great job in bringing a usable order to the vast amounts of web data available. Users are used to rely mostly on the first few results presented in search engine result pages (SERP). Search engines like Google or Bing are constantly optimizing their ranking algorithms to improve the top results on their SERP. While earlier studies on the retrieval effectiveness of web search engines mainly focused on comparing the top results from different search engines, we would like to focus on the comparison of different sections from the same result list. How does the head of a result list compare to it's tail?

In contrast to commercial search engines, so-called long-tail search engines try to support more unfamiliar usage pattern by deliberately redirecting users away from the head of result lists. Prominent examples of such long-tail search engines are bananaslug.com and Million Short. Both search engines incorporate different ideas to access the long tail: While bananaslug.com randomly expands queries with an additional keyword to bring in unexpected results, Million Short removes the most popular websites from the result list. Both search engines point out that these alternative approaches are not meant for everyday use. Their primary goal is to ``offer an interesting view on what usually is hidden from the user" by the ranking algorithms of search engines, the long tail.

Therefore in this study we try to learn more about the long tail of web search engines' result lists. The motivation behind this observation is that web search engine result lists are more diverse than the first one to three hits might suggest \cite{hariri_relevance_2011}. Intuition tells us that the web pages listed on the second, third or even deeper result page might also contain valuable information, but most of the time we don't see them due to the fixation on the top results. By incorporating and reorganising results from the long tail the serendipity effect might be supported. Another motivation might be the wish to explicitly see different results like unpopular, old or controversial web pages that would never be included in the head results due to weak page ranks or other negative ranking factors.

\textbf{Research question.}
Does the long tail as presented by a special long-tail web search engine contain valuable information for users? Can we quantify this using relevance scores gained from an online relevance assessment? In other words: Are the filtered results of a long-tail web search engine better, same or worse compared to a standard result list. What else can we learn about the composition of the long-tail results? 

\textbf{Approach.}
We conducted a relevance assessment study to learn about the relevance distributions in the head and the tail of typical search engine's result lists. We used everyday and domain-specific topics to undertake a relevance assessment study with 33 undergraduate students of a Library and Information Sciences (LIS) course. Each participant judged on up to 30 documents that came from different depths of the long tail of the Million Short web search engine.

\textbf{Related Work.}
Only few studies focus on the analysis of the long tail of web search engines. 
In 2010 Zaragoza et al. reviewed the top five results of 1000 queries sampled from the query log of a major search engine. They report that more than 90\% of these queries are served excellent by all major search engines \cite{zaragoza_web_2010}. Most consequently as reported by Sterling only 8\% of the users are willing to inspect more than three result pages \cite{Sterling2008}. Hariri \cite{hariri_relevance_2011} conducted a study on 34 real search sessions and compared the relevance assessments on the first four result pages (i.e. the first 40 results). While 47.06\% of the first results were judged relevant an even higher percentage of relevant documents (50\%) were found at the 5th SERP position. Even on the 4th results page there were three documents that were judged most relevant by the users in more than 40\% of the searches. In summary Hariri did not find significant differences between the precision of the first four result pages.

\textbf{Contributions.}
While we see a clear difference between the head and the tail of the search engine's result list (measured using Kendall's $\tau$ and intersecting percentages), we see no statistical significant differences in the binary relevance judgments. This means that the tail contains different but still relevant and therefore valuable results for the users. When using graded relevance values we see a slight decrease but still no truly significant difference. Therefore we argue that the long tail contains valuable information and is a rich source for the diversification of web search engine result lists.


\section{Materials and Methods}

In this paper we focus on Million Short, an experimental web search engine that filters the top million (or top 100k, 10k, 1k, 100) sites from the result list. To identify these top sites Million Short is using a combination of its own crawl data and the Alexa Internet traffic statistics. 
To implement the actual retrieval process Million Short is using the Bing API that is augmented with some own crawl data. 
The Million Short website describes the main motivation as: ``We thought it might be somewhat interesting to see what we'd find if we just removed an entire slice of the web"\footnote{\url{https://millionshort.com/about.html}}. 
This slice usually consists of the most popular sites of the web (like Wikipedia, Facebook or Ebay). By removing these sites Million Short pushes results that are in the long tail of the result list (due to low popularity scores, poor search engine optimizations, small marketing budget, non-competitive keywords or simple non-linkage) to the top of it's own results. 
In this paper we will regard the results presented by Million Short as being part of the long tail, although other definitions or implementations are possible. 

The relevance assessments were conducted using a tool called RAT.
The Relevance Assessment Tool (RAT) is a self-developed web-based software that provides an integrated framework for conducting retrieval effectiveness tests on web search engines \cite{lewandowski2013designing}. 
It has been developed to support researchers to design complex system-orientated search engine retrieval effectiveness studies, and to significantly reduce the effort to conduct such tests, as well. By its architecture, it allows us to collect judgements from a multitude of assessors using a crowd-sourcing approach. The assessment tool can be used to design a test, to automatically fetch results from multiple search engines through screen scraping, and to collect relevancy judgments. The toolkit consists of four different modules: (1) test design and project administration, (2) search engine results scraping, (3) assessor interface to the collecting of judgments, and (4) export of assessment results.

The relevance assessments were gathered in a Library and Information Science course called ``Semantics Part II" at Hochschule Darmstadt, Germany in the winter semester of 2012/2013. The task of the assessment exercise was to assess topical relevance (graded relevance) of web pages concerning a given topic. The students of this course were in the second semester and had experiences in evaluating relevance in previous lessons and exercises. In a self assessment they rated their web search experience with an average experience of 7.3 years. The group of assessors consisted of 23 male and 10 female students with an average age of 23.8 years. The users were given an written exercise description with a short oral introduction and a description of the general task and the relevance scale that ranged from 0 (non-relevant) to 4 (fully relevant).  
The topic descriptions were a mixture of specific and broad topics. They covert topics from day-to-day life, celebrities and politics and could be considered as mostly informational and only few navigational or transactional topics. Each topic had a title and a short description that was two to three sentences long. Since we let each assessor evaluate the top 10 results from three different systems for 25 different topics each system delivered 250~results. 

As soon as the assessors logged into the assessment toolkit one of the 25~topics was assigned to them using a round robin approach. After all topics were assigned the following assessors were given a random topic. This resulted in six topics that were rated by more than one assessor. 
Theoretically each assessor had to evaluate 30 single web pages, 10 top results for three different systems.
The three systems are named 0k, 10k, and 1000k. 0k is the Million Short result list without any filtered sites, 10k is the result list with the top 10,000 sites removed and 1000k with the top million sites removed, respectively. In practice due to the pooling process the actual numbers of assessments per topic ranged from 10 to 26. In total we gathered 990 single relevance assessment from 33 different assessors, 30~assessments per assessor. From the total number of 990 assessments we had 459~unique relevance assessments on the websites that formed our pool and were the basis for a clean Qrels file. 
The relevance assessment from different assessors on the same topic were combined using a majority vote approach as described by Hosseini et al. \cite{hosseini_aggregating_2012}. Given a five-point scale we measured inter-rater agreement using Krippendorf's $\alpha$ and found low agreement rates with $\alpha$ values around 0.36. Although the agreement values were generally low and should be handled with care they were in the same range compared to previous studies with non-professional assessors \cite{schaer_better_2012}. We encouraged the assessors to comment on their relevance assessments and gathered 60 free text comments that were manually classified into eight different groups of comments (see Table \ref{tab:comments}).

\section{Results}
\label{sec:results}

\textbf{Differences between 0k, 10k and 1000k and the pooling process.}
We see the different impact the filtering of popular website has on the corresponding result lists per topic. 
When we compare the set-wise intersection between the three systems 0k, 10k and 1000k we see that 0k and 10k share 161 common results while the intersection between 0k and 1000k was only 85 websites. Therefore more than 1/3 of the results from 0k are replaced by long tail results to form 10k and more than 2/3 are replaced to form 1000k (see Table \ref{tab:kendall}). Taking a ranked list-wise and not set-wise look on the results using Kendall's $\tau$ we see no similarities between the different results lists' ranking of 0k, 10k and 1000k. Table \ref{tab:kendall} shows the result of the analysis on Kendall's $\tau$ to check on the consistency between the different systems' rankings. Since all systems values are around 0.1 in average it is clear to say that the three different systems return weak intersecting result sets and non-comparable result lists.
 
\textbf{Retrieval performance.} 
We use two binary (MAP@n and P@n) and one graded (NDCG@n) relevance measure to evaluate the retrieval performance of 0k, 10k, and 1000k. All measures were calculated for n=5 and n=10. The relevance scores are generally very high with a top P@5 value of 0.8 for 0k. 

Although the three systems return different results (in regard to intersections and rankings) the binary performance measures are more or less the same. In fact the differences are so low that we have to compare four decimal places to see an actual difference (i.e. MAP@10 0k: 0.4637 and 1000k: 0.4635). Of course these differences are marginal and therefore not statistical different when applying a Student's t-test. A slightly different situation arises when we interpret the graded relevance levels instead of binary judgements. Here we see a slight drop in NDCG@5 or NDCG@10 that is weakly statistical significant ($\alpha \leq 0.1$).

\begin{table}[t]
\caption{Kendall's $\tau$ (left) and intersection values (right) for all rankings from the three systems 0k, 10k and 1000k.}
\label{tab:kendall}

\parbox{.5\linewidth}{
\centering
\begin{tabular}{r@{\quad}l@{\quad}l@{\quad}l}
\toprule
        & 0k & 10k       &	1000k   \\
\midrule
0k	    & 1 & 0.0984     & 0.0904   \\
10k     &   & 1	         & 0.1160   \\
1000k   &   & 	         & 1        \\
\bottomrule
\end{tabular}}
\hfill
\parbox{.5\linewidth}{
\centering
\begin{tabular}{r@{\quad}l@{\quad}l@{\quad}l}
\toprule
        & 0k    &10k    &1000k      \\
\midrule
0k	    & 250   & 161   & 85        \\
10k     &       & 250	& 130       \\
1000k   &       & 	    & 250       \\
\bottomrule
\end{tabular}}

\bigskip

\centering
\caption{Retrieval results on the three different depths in the long tail. We see no significant differences using a two-tailed t-test with $\alpha \leq 0.05$ but weak significance when using $\alpha \leq 0.1$ (marked with *)}
\label{tab:trec-results}
\begin{tabularx}{\textwidth}{r@{\quad} X X X X X X}
\toprule
	    &   MAP@5   &	MAP@10  &	P@5     &	P@10    &	NDCG@5  &	NDCG@10\\ 
\midrule
0k      &	0.2498	&   0.4637	&   0.8000  &   0.7720	&   0.5845	&   0.6469\\
10k	    &   0.2460	&   0.4647	&   0.7920	&   0.7800  &	0.5625  &	0.6413\\
1000k	&   0.2399	&   0.4635	&   0.7760	&   0.7880	&   0.5413*	&   0.6079*\\
\bottomrule
\end{tabularx}

\bigskip

\caption{Analysis of assessors' comments that could be categorized into eight different groups of comments and their distribution in total and on the three systems.}
\label{tab:comments}
\begin{tabularx}{1\columnwidth}{r@{\quad} X X X X}
\toprule
comment type             & total & 0k & 10k & 1000k \\
\midrule
reliability              & 5     & 0  & 3   & 2     \\
technical error          & 6     & 2  & 2   & 2     \\
language                 & 21    & 8  & 5   & 8     \\
misleading title         & 4     & 2  & 1   & 1     \\
missing content          & 14    & 5  & 4   & 5     \\
paid content             & 2     & 1  & 1   & 0     \\
too specific / too broad & 3     & 0  & 0   & 3     \\
wrong content type       & 20    & 7  & 7   & 6     \\
\bottomrule
\end{tabularx}
\end{table}   

\textbf{Analysis of assessors' comments.}
From the 60 free text comments that were in the data set we extracted eight different types of comments. Each free text was mapped to one or two comment group, depending on the exhaustiveness of the comment (see Table \ref{tab:comments}).  Two comment types have to be highlighted because they only were mentioned for the two long-tail systems 10k and 1000k: reliability and broadness/specificity of the results. The assessors never commented on these two comment types for results from 0k. Given the fact that all other comment types were uniformly distributed between head and tail these two stood out.

\section{Discussion and Conclusion}

We were not able to find significant differences in the retrieval performance of the head and the tail of the Million Short result list for 25 different topics when using MAP@n and P@n. 
The use of graded relevance introduced a slightly different view on the results. We see some weak hints that the retrieval performance of the long tail search engine is not 1:1 comparable to the head. We got some additional hints on differences in the details of the assessors' comments. Analyzing the free text comments we see two types of issues that were only mentioned for the long tail results: reliability and broadness/specificity of the results. To further interpret these results and also other complaints like i.e. language concerns we need more (meta-)data about the actual retrieval sessions. It would be useful to gather these additional data during the scraping process or to allow the integration of additional tools like page classification or language detection systems. Otherwise these data might not be available at a later point. 


We should see the results in the light of the ongoing discussion about the evaluation criterion for IR systems that make a strong argument for having a look at the actual usefulness of the results. Having an evaluation criterion like usefulness might help to better differentiate between the actual characteristics and performance of the long tail compared to the head. This can be seen in the context that we only let our assessors rate on topical relevance while Bing incorporates hundreds of other relevance signals.
A clear limitation of this study is the fact that Million Short is based on Bing which is a black box. 
Nevertheless we see strong hints that our general claim regarding new evaluation methods and (meta-)data for online assessment tools is valid and should be further investigated.

\bibliographystyle{splncs03}
\bibliography{million}

\end{document}